\documentclass[conference,10pt]{IEEEtran}

\ifCLASSINFOpdf

\fi

\usepackage[subrefformat=parens,labelformat=parens,caption=false,font=footnotesize]{subfig}
\usepackage{array}
\usepackage{amsmath}
\usepackage{breqn}
\usepackage[nolist,nohyperlinks]{acronym}
\usepackage{graphicx,wrapfig,lipsum}
\usepackage{rotating}
\usepackage{balance}
\usepackage{amsmath, amssymb, amsthm}
\usepackage{stmaryrd}
\usepackage{tikz}
\usepackage{verbatim}
\usepackage{url}
\usepackage[utf8]{inputenc}
\usepackage[english]{babel}

\usepackage{multicol}
\usepackage{xr-hyper}

\usepackage{scalerel}
\usepackage{multicol}

\usepackage[noadjust]{cite}
\usepackage[normalem]{ulem}
\usepackage{color}
\usepackage{dsfont}
\usepackage{bm}
\usepackage[short]{optidef}
\usepackage{diagbox}
\usepackage{enumitem}
\usepackage{slashbox}
\usepackage[ruled]{algorithm2e}
\usepackage{multirow}
\usepackage[T1]{}
\usepackage{color, colortbl}
\usepackage{babel}
\usepackage{verbatim}
\usepackage{textcomp}
\usepackage{tabulary}
\usepackage{booktabs}
\usepackage{caption}
\usepackage{pgfplots}

\definecolor{Gray}{gray}{0.95}
\definecolor{LightCyan}{rgb}{0.8,0.85,1}
\definecolor{LightBlue}{rgb}{0.6,0.6,1}
\usepackage[font=small]{caption}
\usepgfplotslibrary{fillbetween}

\setlist{nosep}

\usepackage{mathtools}

\begin{document}
\title{Power Consumption Modeling of 5G Multi-Carrier Base Stations: A Machine Learning Approach}

\author{Nicola Piovesan, David L\'opez-P\'erez, Antonio De Domenico, Xinli Geng, Harvey Bao\\
 \small{Huawei Technologies, Paris Research Center, 92100 Boulogne Billancourt, France.} \\
 \small{Email: \{nicola.piovesan, david.lopez.perez, antonio.de.domenico, gengxinli, harvey.baohongqiang\}@huawei.com }}
 
\maketitle
\vspace{-1.7cm}
\thispagestyle{empty}

\begin{abstract}

The fifth generation of the Radio Access Network (RAN) has brought new services, technologies, and paradigms with the corresponding societal benefits. However, the energy consumption of 5G networks is today a concern. In recent years, the design of new methods for decreasing the RAN power consumption has attracted interest from both the research community and standardization bodies, and many energy savings solutions have been proposed. However, there is still a need to understand the power consumption behavior of state-of-the-art base station architectures, such as multi-carrier active antenna units (AAUs), as well as the impact of different network parameters.
In this paper, we present a power consumption model for 5G AAUs based on artificial neural networks. 
We demonstrate that this model achieves good estimation performance, and it is able to capture the benefits of energy saving when dealing with the complexity of multi-carrier base stations architectures. 
Importantly, multiple experiments are carried out to show the advantage of designing a general model able to capture the power consumption behaviors of different types of AAUs. Finally, we provide an analysis of the model scalability and the training data requirements. 
\end{abstract}

%%%%%%%%%%%%%%%%%%%%%%%%%%%%%%%%%%%%%%%%%%%%%%%%%%%%%
\section{Introduction}
\label{sec:intro}
%%%%%%%%%%%%%%%%%%%%%%%%%%%%%%%%%%%%%%%%%%%%%%%%%%%%%
Recent studies indicate that, 
by 2030, 
the number of connected devices is expected to increase to 100 billion, 
and that \ac{5G} mobile networks may be supporting up to 1,000× more data traffic than the \ac{4G} ones in 2018.
However, the energy consumption of future networks is concerning. 
Deployed \ac{5G} networks have been estimated to be approximately four times more energy efficient than 4G ones. 
Nevertheless, their energy consumption is around three times higher, 
due to the larger number of cells required to provide the same coverage at higher frequencies, 
and the increased processing required by their larger bandwidths and more antennas~\cite{Huawei2020}.
It should be noted that, 
on average, 
the network \ac{OPEX} accounts for approximately 25\% of the total costs incurred by a \ac{MNO}, 
and that 90\% of it is spent on large energy bills~\cite{GSMA20205Genergy}. 
Importantly, 
more than 70\% of this energy has been estimated to be consumed by the \ac{RAN}, 
and in more details,
by the \acp{BS}~\cite{lopezperez2022survey}.

The energy challenge of \acp{MNO} is thus to meet the upcoming more challenging traffic demands and requirements with significantly less energy consumption and \ac{GHG} emissions than today to reduce the environmental impact of mobile networks, 
and in turn, costs.

Third generation partnership project (3GPP) new radio (NR) Release 15 specified intra-NR network energy saving solutions 
--similar to those developed for 3GPP long-term evolution (LTE)-- 
to decrease RAN energy consumption. 
Moreover, 3GPP NR Release 17 has recently specified inter-system network energy saving solutions, 
and is currently taking network energy saving as an artificial intelligence use case.
However, data collected from 3GPP LTE and NR networks have shown that these solutions are still not sufficient to fundamentally address the challenge of reducing energy consumption~\cite{CT2021report}. 
For this reason, 
3GPP NR Release 18 has recently approached a study item, 
which attempts to develop a set of flexible and dynamic network energy saving solutions~\cite{CT2021report}.
Importantly, 
this study item indicates that new 5G power consumption models are needed to accurately develop and optimize new energy saving solutions, 
while also considering the complexity emerging from the implementation of state-of-the-art base station architectures.

In recent years, 
many models for base station power consumption have been proposed in the literature.
The work in~\cite{Auer2011} proposed a widely used power consumption model, 
which explicitly shows the linear relationship between the power transmitted by the BS and its consumed power. 
This model was extended in \cite{Debaillie2015}, 
taking into account the \ac{mMIMO} architecture and energy saving methods. 
However, the power consumption estimate discussed in that paper seems to be inaccurate~\cite{han2020energy} with an optimistic 40.5~W per \ac{mMIMO} BS.
The work in~\cite{Tombaz2015} further extended the model in~\cite{Auer2011} by considering a linear increase of the power consumption with the number of mMIMO transceivers. 
A more complete and detailed description of the power consumption components was introduced in~\cite{bjornson2015optimal}, 
where the authors provided a model that considers the mMIMO architecture, downlink and uplink communication phases, as well as the number of multiplexed users per \ac{PRB}, 
and a large number of mMIMO components. 
The power consumption of a system that uses multiple carriers was modeled in~\cite{Yu2015} by considering a linerar model. 
Finally, the work in~\cite{lopez2021energy} jointly considered mMIMO and multi-carrier capabilities, 
such as carrier aggregation and its different aggregation capabilities.

Aiming at providing more accurate estimations,
validated in the field,
in our most recent work~\cite{piovesan2022machine}, 
we introduced a new analytical power consumption model for 5G \acp{AAU}
-- the highest power-consuming component of today's mobile networks, based on a \ac{ML} framework, which builds on a large data collection campaign.
In this paper, 
we present in detail our \ac{ML} framework
providing a detailed technical analysis of its accuracy performance, its scalability, and generalization capabilities.

%%%%%%%%%%%%%%%%%%%%%%%%%%%%%%%%%%%%%%%%%%%%%%%%%%%%%
\section{5G AAU architecture}
\label{sec:AAUmodel}
%%%%%%%%%%%%%%%%%%%%%%%%%%%%%%%%%%%%%%%%%%%%%%%%%%%%%

\begin{figure}
    \centering
    \includegraphics[scale=0.8]{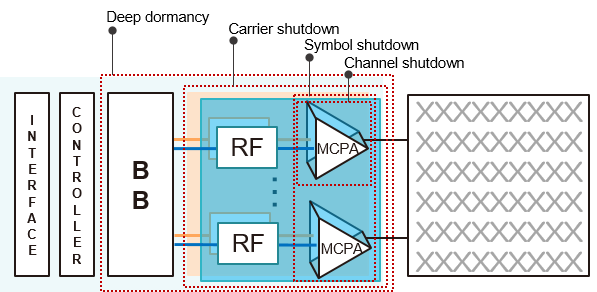}
    \caption{Architecture of an AAU with MCPAs handling 2 carriers in 2 different bands, which transmit over the same wideband antenna panel.}
    \label{fig:AAUarchitecture}
\end{figure}

The hardware architecture of a 5G \ac{AAU} is shown in Fig.~\ref{fig:AAUarchitecture}. 
In particular, 
in our \ac{AAU} architecture, 
we assume that:
\begin{itemize}
\item 
The AAU has a multi-carrier structure, 
and uses \ac{MCPA} technology;
\item 
The \ac{AAU} manages $C$ carriers deployed in $T$ different frequency bands;
\item 
The \ac{AAU} comprises $T$ transceivers, 
each operating a different frequency band, 
and $M$ MCPAs, 
one for each antenna port;
\item 
A transceiver includes $M$ \ac{RF} chains, 
one per antenna port, 
which comprise a cascade of hardware components for analog signal processing, 
such as filters and digital-to-analog converters;
\item 
Antenna elements are passive. 
For example, one wideband panel or $T$ antenna panels may be used per \ac{AAU};
\item 
Deep dormancy, carrier shutdown, channel shutdown, and symbol shutdown are implemented, 
each switching off distinct components of the \ac{AAU} (as shown in Fig.~\ref{fig:AAUarchitecture}). 
\end{itemize}

Importantly, 
it should be noted that the implementation of \acp{MCPA} leads to increased energy efficiency compared to single carriers \acp{PA}, 
as the management of multiple carriers through one `wider' \ac{PA} allows to manage a larger amount of transmit power, 
in turn, permitting the \acp{MCPA} to operate at higher energy efficiency areas. 
Moreover, the static power consumption of the \acp{MCPA} increases sub-linearly with the number of carriers, 
since part of the hardware components can be shared among them.
However, it should be noted that the implementation of \acp{MCPA} involves increased complexity in the management of the network energy saving methods and in the estimation of the power consumption when such methods are activated. 
In fact, the deactivation of one carrier may not bring the expected energy savings, 
if the \acp{MCPA} need to remain active to operate the co-deployed carriers.

%%%%%%%%%%%%%%%%%%%%%%%%%%%%%%%%%%%%%%%%%%%%%%%%%%%%%
\section{ANN Model Architecture}
\label{sec:model}
%%%%%%%%%%%%%%%%%%%%%%%%%%%%%%%%%%%%%%%%%%%%%%%%%%%%%

In this section, 
we describe the data collected during our measurement campaign, 
and we provide a description of the ANN architecture designed for modeling and estimating the power consumption. 
Moreover, we describe the identified loss function and the training of the ANN model parameters.

\subsection{Dataset}

The dataset used for training and testing the \ac{ANN} model is composed of hourly measurements collected during 12 days from a deployment of 7500 \ac{4G}/\ac{5G} \acp{AAU}. 
Overall, 24 different types of \acp{AAU} are included. 
The collected measurements contain 150 different features,
which can be classified into four main categories:
\begin{itemize}
    \item \textit{Engineering parameters}: 
    Information related to the configuration of each \ac{AAU} 
    (e.g., \ac{AAU} type, number of \acp{TRX}, numbers of supported and configured carriers);
    \item \textit{Traffic statistics}: 
    Information on the serviced traffic 
    (e.g., average number of active \acp{UE} per \ac{TTI}, number of used \acp{PRB}, traffic volume serviced);
    \item \textit{Energy saving statistics}: 
    Information on activated energy saving modes \cite{lopezperez2022survey}
    (e.g., duration of the carrier-shutdown, channel-shutdown, symbol shutdown and dormancy activation);
    \item \textit{Power consumption statistics}: 
    Information on the power consumed by the \ac{AAU}.
\end{itemize}

\begin{figure}
    \centering
    \includegraphics[scale=0.65]{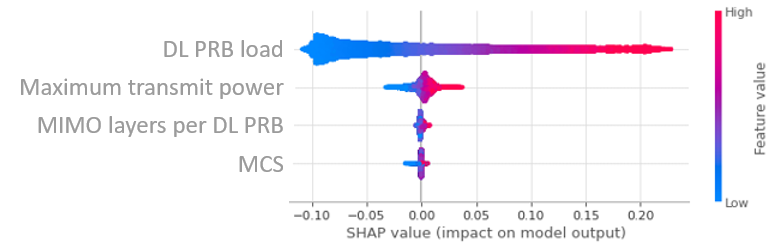}
    \caption{Example of SHAP analysis performed on 4 of the available features is the collected measurements data.}
    \label{fig:shap}
\end{figure}

Feature importance analysis has been extensively performed on the collected features to identify the most relevant for estimating power consumption. 
It is worth highlighting that the features that do not affect power consumption as well as those highly correlated with the selected ones (i.e., thus providing limited information) were discarded.
The analysis of the feature importance consisted into two phases: 
i) a first phase, 
in which gradient boosting models including different input features were trained, and 
ii) a second phase,
in which the analysis of \ac{SHAP} values~\cite{NIPS2017_7062} were performed on such models. 
The \ac{SHAP} value of each feature represents the change in the expected model prediction when conditioning on that feature.
 
As an example, 
Fig.~\ref{fig:shap} shows the \ac{SHAP} values of four features, 
namely the \ac{DL} \ac{PRB} load, the maximum transmit power, the number of \ac{MIMO} layers per \ac{PRB}, and the \ac{MCS}. 
In more details,
the figure indicates in which direction and how much each feature contributes to the model output as compared to the average model prediction. 
The y-axis on the right side indicates the respective feature value (low values in blue color and high values in red). 
Each scatter dot represents one instance in the data.

The analysis highlights that the \ac{DL} PRB load is the most important feature, 
whereas the maximum transmit power is the second most important. 
In fact, the knowledge of this two features allows the model to capture the amount of power transmitted by the AAU at different \ac{DL} \ac{PRB} load levels. 
In more detail, 
the model output is shown to increase when the DL PRB load and/or the maximum transmit power are increased.

Importantly, 
the \ac{MCS} and the number of \ac{MIMO} layers per \ac{DL} \ac{PRB} show a large correlation with the \ac{DL} \ac{PRB} load, 
meaning that the latter feature is sufficient to capture the energy consumption behavior. 
The extended analysis of the importance of the available features allowed us to identify the inputs needed for our \ac{ANN} model, which correspond to the type of \ac{AAU} and a set of characteristics for each of the carriers. 
The complete list of selected features is shown in Table~\ref{tab:ML_inputs}.

\begin{table}[]
\begin{tabular}{@{}lll@{}}
\toprule
Class                    & Parameter                        & Type               \\ \midrule
Engineering parameter    & AAU type                         & Categorical        \\
Engineering parameter    & Number of  TRXs                  & Numerical          \\
Engineering parameter    & Carrier transmission mode        & Categorical \\
Engineering parameter    & Carrier frequency                & Numerical          \\
Engineering parameter    & Carrier bandwidth                & Numerical          \\
Engineering parameter    & Carrier maximum transmit power   & Numerical \\
Traffic statistics       & Carrier DL PRB load              & Numerical          \\
Energy saving statistics & Duration of carrier shutdown     & Numerical          \\
Energy saving statistics & Duration of channel shutdown     & Numerical          \\
Energy saving statistics & Duration of symbol shutdown      & Numerical          \\
Energy saving statistics & Duration of deep dormancy        & Numerical          \\ \bottomrule
\end{tabular}
\caption{ANN model input parameters.}
\label{tab:ML_inputs}
\end{table}

\subsection{Inputs of the model}
\label{sec:inputlayer}

Each of the input features listed in Table~\ref{tab:ML_inputs} was pre-processed according to its type, 
and inputted to the neural network. 
The numerical features were normalized before being inputted into the model, 
whereas the categorical ones were inputted by using one-hot encoding.

Since a \ac{AAU} can operate multiple carriers through the same \acp{MCPA},
to make our \ac{ANN} model the most general and flexible,
it takes input data from $C^{\mathrm{MAX}}$ carriers, 
which corresponds to the maximum number of carriers that can be managed by the most capable \ac{AAU} in the dataset.
When $C<C^{\mathrm{MAX}}$ carriers are deployed in an \ac{AAU}, 
all the input neurons related to the remaining $C^{\mathrm{MAX}}-C$ carriers are set to zero. 
It is worth noting that this approach allows to implement an \ac{ANN} model with a fixed number of input neurons,
which can be trained with data from all the AAUs in the dataset, 
regardless of their number of configured carriers, 
with a minimal loss in terms of accuracy, 
as we will discussed in Section~\ref{sec:zeros}. 

\subsection{Outputs of the model}

The analysis of the collected data has highlighted that different power consumption values may be reported for the same input feature values. 
This effect have multiple origins: 
i) the presence of features slightly impacting the power consumption but currently not captured as input for the model, 
ii) errors in the measurements or in the collection of the data, 
iii) tolerance of the hardware components which affects their power consumption behavior. 

To embrace such noise,
we define the measured power consumption, $\bar{y}$, as $\bar{y} = y + n$, 
where 
$y$ is the power consumption for a given input configuration,
and $n$ is the noise due to the mentioned errors.
Based on the analysis of the available data, 
the noise, $n$, can be assumed to be normally distributed with mean 0 and standard deviation $\sigma$. 
It thus follows that the measured power consumption, $\bar{y}$, is normally distributed with mean $\mu=\mathbb{E}[y]$ and standard deviation $\sigma$. 

The designed \ac{ANN} model estimates and outputs these two parameters, $\mu$ and $\sigma$. 
Furthermore, it is worth highlighting that the output of these two parameters also allows computing a confidence interval for each power consumption estimate, 
thus increasing the reliability of the whole process.

\subsection{Architecture of the model}
\begin{figure}
    \centering
    \includegraphics[scale=0.6]{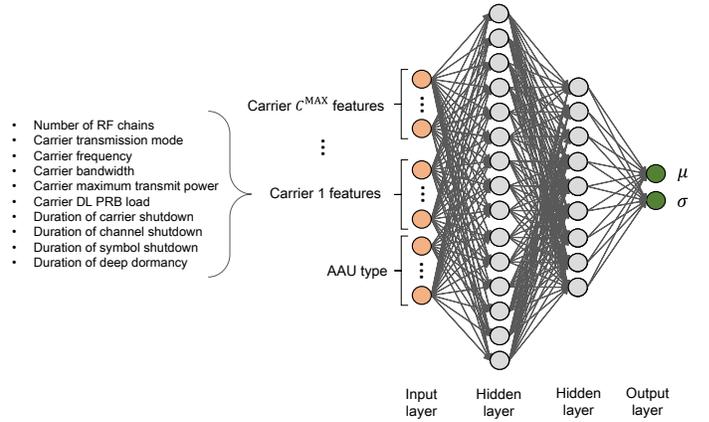}
    \caption{Architecture of the ANN.}
    \label{fig:MLmodel}
\end{figure}

Multilayer perceptron is considered as the basic architecture for the proposed \ac{ANN} model, 
consisting of multiple fully connected layers of neurons~\cite{Goodfellow-et-al-2016}. 
The overall structure of the proposed \ac{ANN} model is depicted in Fig.~\ref{fig:MLmodel}.

In general, 
the input layer is comprised of $n_i=N_{\mathrm{AAU}}+10*C^{\mathrm{MAX}}$ neurons, 
where $N_{\mathrm{AAU}}$ is the number of \ac{AAU} types available in the dataset, 
and thus modelled by the \ac{AAU},
and $C^{\mathrm{MAX}}$ is the maximum number of carriers of the most capable \ac{AAU},
as discussed earlier. 
In our specific scenario, 
we collected data for $N_{\mathrm{AAU}}=24$ different \ac{AAU} types,
and the maximum number of carriers of the most capable AAU, $C^{\mathrm{MAX}}$, is equal to 6.
Therefore, the input layer consists of $n_i=84$ neurons.

The input layer is followed by two hidden layers,
which are composed of 40 and 15 neurons, respectively. 
These dimensions were chosen after an optimization process aimed at maximizing the accuracy of the model. 

Finally, the output layer is composed of two neurons, 
which capture the mean and standard deviation of the power consumption,
as explained earlier.
As both metrics must be positive, 
the sigmoid activation function is adopted at the output layer.

\subsection{Training of the model}

The goal of the model optimization process is to minimize both the prediction error and the uncertainty. 
More in detail, 
the \ac{ANN} training process is considered successful if the statistical distribution of the power measurements outputted by the model for a given input, $x$, matches the distribution of the power measurements in the data. 
Therefore, during the training phase, 
the aim is to maximize the probability that the power consumption estimates, $\bar{y}$, belong to ---are within--- the distribution $\mathcal{N}(\mu,\sigma)$. 

Since the power consumption, $\bar{y}$, follows a normal distribution, 
this probability is computed as
\begin{equation}
    P\left(\bar{y}|\mu,\sigma\right) = \frac{1}{\sigma\sqrt{2\pi}} e^{-\frac{(\bar{y}-\mu)^2}{2\sigma^2}}.
\end{equation}

As most of the optimizers used to train \acp{ANN} are designed to solve minimization problems, 
we consider the following loss function to train the ANN model:
\begin{equation}
    l(\bar{y},\mu,\sigma) = - \log \left( P(\bar{y}|\mu,\sigma)\right) = \log(\sigma) + \frac{(\bar{y}-\mu)^2}{2\sigma^2}.
    \label{eq:loss}
\end{equation}

It should be noted that this function reflects the goal of reducing both the prediction error and the related uncertainty.
In fact, 
the first term is minimized when the standard deviation, $\sigma$, is low,
which means that the confidence in the estimation is high, 
whereas the second term is minimized when the prediction error, $\bar{y}-\mu$, is reduced. 

Before the model training was performed, 
the available data set was split into two parts: 
a training set and a testing set. 
The training set contains data collected for 10 days from our 7500 \acp{AAU}, 
whereas the testing set contains data collected for 2 days from the same \acp{AAU}. 
In addition, 
80,\% of the training samples are randomly selected to train the \ac{AAU} model, 
whereas the remaining 20\,\% are used for validating the model during the training phase.

Model training was carried out by adopting the Adam version of the gradient descent algorithm~\cite{Goodfellow-et-al-2016}, 
and required 75 minutes to perform 1086 iterations when adopting a learning rate $\alpha=0.001$. 
Note that an early stopping method was implemented to stop the training after 200 epochs with no improvements in terms of validation loss. 

\section{Experiments and Analysis}

In this section, 
we provide an analysis of the error performance achieved by the ANN model. 
Moreover, we present a set of experiments carried out to evaluate the generalization capabilities of the framework and its scalability related to multi-carrier architectures and AAU types. 
Finally, we investigate the impact of data availability to the estimation performance.

\subsection{Overall model performance}

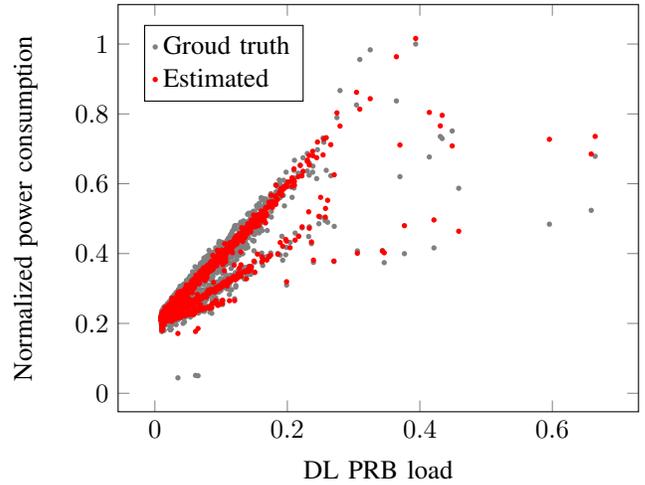
\begin{figure}
    \centering
    \begin{tikzpicture}
    \begin{axis}[xlabel={DL PRB load}, ylabel={Normalized power consumption},width=8.5cm,height=7cm, legend style={at={(0.05,0.95)},anchor=north west},legend cell align={left}]
    \addplot[only marks, mark size=1pt, mark options={fill=gray,draw opacity=0}] table [x=x, y=r, col sep=comma] {Figures/plot.csv};
    \addlegendentry{Groud truth}
    \addplot[only marks, mark size=1pt,mark options={fill=red,draw opacity=0}] table [x=x, y=e, col sep=comma] {Figures/plot.csv};
    \addlegendentry{Estimated}
    \end{axis}
    \end{tikzpicture}
    \caption{True and estimated normalized power consumption vs DL PRB load for multiple BSs of a given type.}
    \label{fig:truevsest}
\end{figure}

To assess the performance of the developed framework, 
we adopted the ANN model to estimate the power consumption of all the 7500 AAUs over the two testing days.
Then, we compared the estimated power consumption with the real measurements available in the data.

In this paper, 
we adopt the \ac{MAE} as a metric to measure the absolute error, 
and the \ac{MAPE} as a metric to evaluate the relative error.
Overall, the model achieved a \ac{MAE} of 10.94~W and a remarkably low \ac{MAPE} of 5.87~\% when estimating the power
consumed by each AAU across all hours of the test period.
As an example, Fig.~\ref{fig:truevsest} shows the real and estimated normalized power consumption for multiple AAUs of the same type. Note that the power consumption linearly increases with the DL PRB load and that three different slopes are observed due to the presence of three different configuration of the maximum transmit power. Notice that the proposed ANN model accurately fits the power consumption for each of the three configurations.

\subsection{Multi-carrier generalization capabilities}
\label{sec:zeros}

As mentioned in Section~\ref{sec:inputlayer}, 
to make the ANN model general and work with any type of AAU, 
the input layer is designed to take input data from $C^{\mathrm{MAX}}$ carriers. 
When $C<C^{\mathrm{MAX}}$ carriers are deployed in the AAU, 
all input neurons related to the remaining $C^{\mathrm{MAX}}-C$ carriers are set to zero.
It is worth noting that the alternative modeling approach consists in training multiple ANN models, 
each of them supporting AAUs with a given number of carriers.
In this section, 
we evaluate the performance loss due to such a general implementation of the \ac{ANN} model.
The performance analysis is performed by considering the following two models:
\begin{itemize}
    \item \textit{Single-carrier ANN model}: 
    The model is tailored to \acp{AAU} in which a single-carrier is deployed 
    (i.e., the input layer is composed of 34 neurons), 
    and is thus exclusively trained with data collected from such \acp{AAU}; 
    \item \textit{General ANN model}: 
    The model provides power consumption estimation for \acp{AAU} with up to $C^{\mathrm{MAX}}=6$ configured carriers 
    (i.e., the input layer is composed of 84 neurons), 
    and is trained with all available AAUs.
\end{itemize}
These two models have been tested to estimate the power consumption of all single-carrier \acp{AAU} available in the collected data. 
In such single-carrier \acp{AAU} test, 
the single-carrier ANN model achieves \ac{MAE} 10.11\,W, and \ac{MAPE} 6.42\,\%, 
whereas the general ANN model achieves \ac{MAE} 10.25\,W and \ac{MAPE} 6.54\,\%.

It should be noted that this performance is different than that presented in the previous section as here we estimate the power consumption only of the single carrier AAUs in our dataset. Also, we can observe that general ANN model achieves slightly worse performance 
(1.38\% loss in terms of MAE and 1.87\% loss in terms of MAPE), 
as it is trained over a more heterogeneous set of data, 
while also needing to capture the complex power consumption behaviors that emerge when considering multi-carrier architectures.
However, these errors are minimal,
and shows that the devised general model can cope with a wider set of \acp{AAU} at the expense of a small cost in terms of performance loss.
Importantly, 
it is worth stressing that the general \ac{ANN} model has the advantage of observing how power consumption depends on multiple input features in a wide variety of \ac{AAU} types, 
and thus, 
as we will see in the next section, 
it can generalize among them. 

\subsection{AAU type generalization capabilities}

In this section, 
we analyze the capability of the ANN power consumption model of generalizing over multiple types of AAU. 
In this way, 
we want to highlight the advantage of our modeling approach, 
in which a single and general model is used to capture the power consumption of large variety of AAU type and configurations.

To assess such capability, 
we select the most popular AAU type in our data, 
and we evaluate the generalisation capabilities of the designed framework by analysing the following models:
\begin{itemize}
    \item \textit{Single-AAU ANN model}:
    The model is trained --and can provide power estimations-- exclusively for the selected AAU type. 
    Moreover, the training data does not include any sample in which carrier shutdown is activated.
    \item \textit{General ANN model}:
    The model is trained with data collected by all the AAUs. 
    As in the previous case, 
    the training data related to the selected AAU does not include any sample in which carrier shutdown is activated. 
    However, training data related to other types of AAUs includes samples in which the carrier shutdown feature is activated. 
\end{itemize}
The two models are tested to estimate the power consumption of the selected AAU over the testing set, 
in which carrier shutdown is activated for some periods.
The single-AAU ANN model leads to poor accuracy estimations 
(i.e., MAE 57.82\,W, MAPE 10.04\,\%),
as it is not able to learn how to characterise the carrier shutdown feature due to the poor training data. 
However, the general ANN model provides improved performance 
(i.e., \ac{MAE} 19.32\,W, \ac{MAPE} 3.59\,\%),
even if there is not training data covering the scenario in which carrier shutdown is activated for the selected \ac{AAU}. 

We highlight that such improved performance is achieved thanks to the generalization capability of our \ac{ANN} model, 
which allows capturing knowledge from many different types of \ac{AAU}. 

\subsection{ANN scalability}

\begin{figure}
    \centering
    \begin{tikzpicture}
    \begin{axis}[xlabel={$N$}, ylabel={MAPE [\%]},width=8.5cm,height=5cm,grid=both, legend style={at={(0.05,0.95)},anchor=north west},legend cell align={left}]
    \addplot[mark=*] coordinates {
        (5, 5.914)
        (7, 6.020)
        (10, 6.1306)
        (12, 6.2924)
        (15,6.4545)
        (17, 6.538)
        (20,6.655)
    };
    \addlegendentry{fixed $c=1$}
    \addplot[mark=none,color=red,style=dashed,mark=square*, mark options={solid}] coordinates {
        (5, 5.914)
        (7, 5.850)
        (10,5.875)
        (12,5.971)
        (15,5.928)
        (17,5.924)
        (20,5.870)
    };
    \addlegendentry{scalable $c$}
   
    \end{axis}
    \end{tikzpicture}
    \caption{MAPE achieved by the ANN model when trained/tested over $N$ AAUs types.}
    \label{fig:performancePerProduct}
\end{figure}
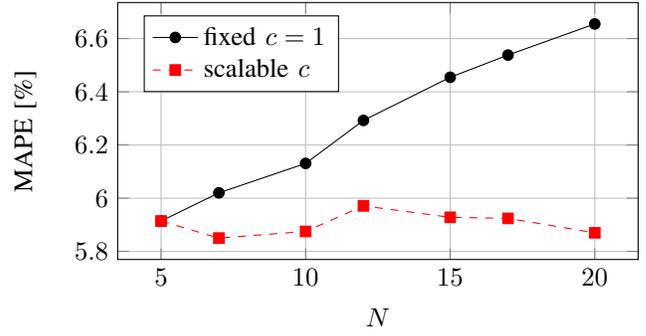

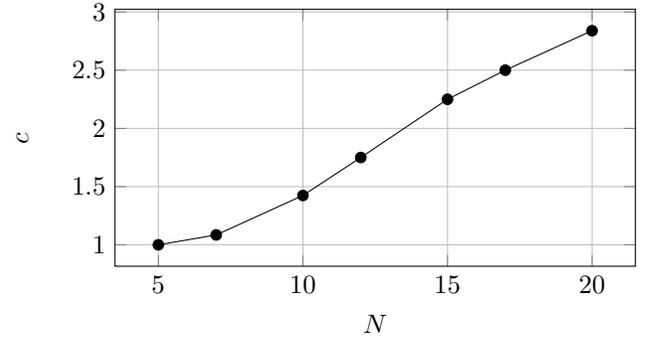
\begin{figure}
    \centering
    \begin{tikzpicture}
    \begin{axis}[xlabel={$N$}, ylabel={$c$},width=8.5cm,height=5cm,grid=both]
    \addplot[mark=*] coordinates {
        (5, 1)
        (7,1.085)
        (10,1.424)
        (12,1.75)
        (15,2.25)
        (17,2.5)
        (20,2.839)
    };

    \end{axis}
    \end{tikzpicture}
    \caption{ANN scaling factor, $c$, adopted for different number of AAU types included in the data, $N$.}
    \label{fig:performanceScaling}
\end{figure}

The results discussed in the previous sections highlight that the proposed framework is capable of providing accurate estimations of power consumption when dealing with the complexity of multi-carrier \ac{AAU} architectures. 
Importantly, the model is capable of capturing the power consumption behaviors of each AAU type considering 5G energy saving features. 
In this section, 
we analyze how the dimension of the ANN architecture should be scaled according to the number of \ac{AAU} types included in the data. 

As a starting scenario, 
we consider a dataset that includes 5 different types of \ac{AAU}. 
Multiple \ac{ANN} shapes/sizes were trained and tested to identify the smallest \ac{ANN} providing a good estimation error. 
The identified \ac{ANN} is composed of two hidden layers with, respectively, $l_1=12$ and $l_2=4$ neurons, 
and it reaches \ac{MAE} 11.02 W and \ac{MAPE} 5.91\%.

The same \ac{ANN} architecture was trained and tested on datasets including a larger number of \ac{AAU} types. 
Fig.~\ref{fig:performancePerProduct} shows in black the performance achieved when increasing the number of \ac{AAU} types in the dataset.
It can be seen that the estimation error deteriorates when increasing the number of \ac{AAU} types in the data. 
In particular, 
the \ac{MAPE} increases by 3.6\%, 9.1\% and 12.4\% when increasing the number of \ac{AAU} types to 10, 15 and 20, respectively.
This error performance degradation is motivated by the fact that, 
when increasing the number of \ac{AAU} types in the data, 
the dimension of the \ac{ANN} architecture (i.e., the parameters of the model) is no longer sufficient to gather the knowledge of the power consumption behavior of different AAU types that allow to successfully estimate their power consumption.

Therefore, the dimension of the \ac{ANN} must be properly scaled when increasing the number of \ac{AAU} types in the data. 
To visualize this, 
we consider a scaling factor for the \ac{ANN} architecture, $c$. 
In more detail, 
the first hidden layer has dimension $l_1=12\cdot c$, 
while the second has dimension $l_2=4\cdot c$.

Different values of the scaling factor $c$ were tested when considering different numbers of \ac{AAU} types in the data,
to assess how the \ac{ANN} should be scaled to guarantee the \ac{MAPE} of the estimation to be within 1\% of the initial error of 5.91\%. 
Fig.~\ref{fig:performanceScaling} shows the lowest value of the scaling factor $c$ that allows to meet the requirement for each number of \ac{AAU} types in the data. 
The results indicates that linearly scaling the dimension of the \ac{ANN} allows us to preserve the accuracy of the ANN estimation, while increasing the number of \ac{AAU} types that need to be modeled.

\subsection{Training data availability}

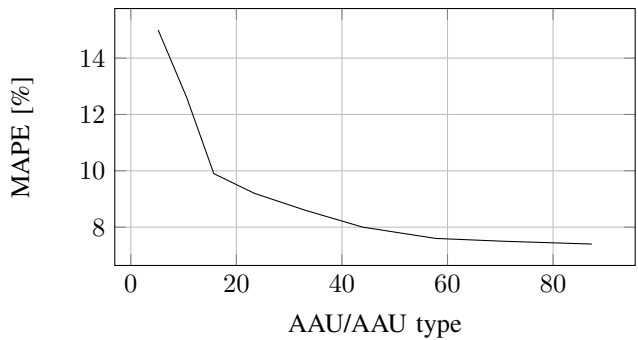
\begin{figure}
    \centering
    \begin{tikzpicture}
    \begin{axis}[xlabel={AAU/AAU type}, ylabel={MAPE [\%]},width=8.5cm,height=5cm,grid=both]
    \addplot[mark=none] coordinates {
        (5.178, 15.0)
        (10.594, 12.6)
        (15.724, 9.9)
        (23.420, 9.2)
        (33.040, 8.6)
        (43.943, 8.0)
        (57.767, 7.6)
        (70.523, 7.5)
        (87.340, 7.4)
    };
    \end{axis}
    \end{tikzpicture}
    \caption{MAPE achieved by the ANN model when considering different number of AAUs per AAU type in the training dataset.}
    \label{fig:AAUproduct}
\end{figure}

Collecting measurements from large network deployments can be challenging and time-consuming.
In this section, 
we analyze how the amount of available training data affects the performance achieved by the \ac{ANN} model. 
To perform the analysis, 
we focus on 11~\ac{AAU} types for which more than 90~\acp{AAU} per type are available in the collected data.

During the training of the ANN model, 
a varying number of AAUs was included for each AAU type, 
while during the testing phase, 
all the available AAUs were considered.

Fig.~\ref{fig:AAUproduct} shows the MAPE achieved by the ANN model when considering a different number of AAUs per AAU type in the training dataset. 
The error achieved has a clear decreasing trend, 
suggesting that including more \ac{AAU} in the training set is beneficial to improve the accuracy of the estimation. 
However, after reaching 70~\ac{AAU} per AAU type, 
adding more \ac{AAU} in the training data provides negligible gains (i.e., lower than 1\%).
\balance
%%%%%%%%%%%%%%%%%%%%%%%%%%%%%%%%%%%%%%%%%%%%%%%%%%%%%
\section{Conclusions}
\label{sec:conclusions}
%%%%%%%%%%%%%%%%%%%%%%%%%%%%%%%%%%%%%%%%%%%%%%%%%%%%%
In this paper, 
we presented a power consumption model for 5G AAUs based on an ANN architecture. 
The ANN model was trained with data collected from a large deployment,
which includes multiple types of AAU with different configurations. 
Feature analysis allowed us to identify a set of input features for the model.
The analysis of the results highlighted that the model can achieve high accuracy, 
with a MAPE less than 6\% when tested on all available AAUs in our data. 
Moreover, the experiments highlighted the advantage of training a single general model over all the AAUs in the data, 
which is able to capture and generalize the impact of multiple parameters on the power consumption and the benefit of energy saving schemes in complex multi-carrier architectures. 
Importantly, the results provided good insights into how the ANN architecture should be scaled when needed to model more AAU types. Moreover, experiments showed that at least 70 AAUs per type should be included in the training to guarantee the achievement of good error performance.

%%%%%%%%%%%%%%%%%%%%%%%%%%%%%%%%%%%%%%%%%%%%%%%%%%%%%
%\vspace{-0.2cm}
\bibliographystyle{IEEEtran}
\bibliography{reference.bib}

\begin{acronym}[AAAAAAAAA]

 \acro{2D}{two-dimensional}
 \acro{3D}{three-dimensional}
 \acro{3G}{third generation}
 \acro{3GPP}{third generation partnership project}
 \acro{4G}{fourth generation}
 \acro{5G}{fifth generation}
 \acro{5GC}{5G Core Network}
 \acro{AAA}{authentication, authorisation and accounting}
 \acro{AAU}{active antenna unit}
 \acro{ABS}{almost blank subframe}
 \acro{AC}{alternating current}
 \acro{ACIR}{adjacent channel interference rejection ratio}
 \acro{ACK}{acknowledgment}
 \acro{ACL}{allowed CSG list}
 \acro{ACLR}{adjacent channel leakage ratio}
 \acro{ACPR}{adjacent channel power ratio}
 \acro{ACS}{adjacent channel selectivity}
 \acro{ADC}{analog-to-digital converter}
 \acro{ADSL}{asymmetric digital subscriber line}
 \acro{AEE}{area energy efficiency}
 \acro{AF}{amplify-and-forward}
 \acro{AGCH}{access grant channel}
 \acro{AGG}{aggressor cell}
 \acro{AH}{authentication header}
 \acro{AI}{artificial intelligence}
 \acro{AKA}{authentication and key agreement}
 \acro{AMC}{adaptive modulation and coding}
 \acro{ANN}{artificial neural network}
 \acro{ANR}{automatic neighbour relation}
 \acro{AoA}{angle of arrival}
 \acro{AoD}{angle of departure}
 \acro{APC}{area power consumption}
 \acro{API}{application programming interface}
 \acro{APP}{a posteriori probability}
 \acro{AR}{augmented reality}
 \acro{ARIMA}{autoregressive integrated moving average}
 \acro{ARQ}{automatic repeat request}
 \acro{AS}{access stratum}
 \acro{ASE}{area spectral efficiency}
 \acro{ASM}{advanced sleep mode}
 \acro{ASN}{access service network}
 \acro{ATM}{asynchronous transfer mode}
 \acro{ATSC}{Advanced Television Systems Committee}
 \acro{AUC}{authentication centre}
 \acro{AWGN}{additive white gaussian noise}
 \acro{BB}{baseband}
 \acro{BBU}{baseband unit}
 \acro{BCCH}{broadcast control channel}
 \acro{BCH}{broadcast channel}
 \acro{BCJR}{Bahl-Cocke-Jelinek-Raviv} 
 \acro{BE}{best effort}
 \acro{BER}{bit error rate}
 \acro{BLER}{block error rate}
 \acro{BPSK}{binary phase-shift keying}
 \acro{BR}{bit rate}
 \acro{BS}{base station}
 \acro{BSC}{base station controller}
 \acro{BSIC}{base station identity code}
 \acro{BSP}{binary space partitioning}
 \acro{BSS}{blind source separation}
 \acro{BTS}{base transceiver station}
 \acro{BWP}{Bandwidth Part}
 \acro{CA}{carrier aggregation}
 \acro{CAC}{call admission control}
 \acro{CaCo}{carrier component}
 \acro{CAPEX}{capital expenditure}
 \acro{capex}{capital expenses}
 \acro{CAS}{cluster angular spread}
 \acro{CATV}{community antenna television}
 \acro{CAZAC}{constant amplitude zero auto-correlation}
 \acro{CC}{component carrier}
 \acro{CCCH}{common control channel}
 \acro{CCDF}{complementary cumulative distribution function}
 \acro{CCE}{control channel element}
 \acro{CCO}{coverage and capacity optimisation}
 \acro{CCPCH}{common control physical channel}
 \acro{CCRS}{coordinated and cooperative relay system}
 \acro{CCTrCH}{coded composite transport channel}
 \acro{CDF}{cumulative distribution function}
 \acro{CDMA}{code division multiple access}
 \acro{CDS}{cluster delay spread}
 \acro{CESM}{capacity effective SINR mapping}
 \acro{CO$_{2e}$}{carbon dioxide equivalent}
 \acro{CFI}{control format indicator}
 \acro{CFL}{Courant-Friedrichs-Lewy}
 \acro{CGI}{cell global identity}
 \acro{CID}{connection identifier}
 \acro{CIF}{carrier indicator field}
 \acro{CIO}{cell individual offset}
 \acro{CIR}{channel impulse response}
 \acro{CNN}{Convolutional Neural Network}
 \acro{CMF}{cumulative mass function}
 \acro{C-MIMO}{cooperative MIMO}
 \acro{CN}{core network}
 \acro{COC}{cell outage compensation}
 \acro{COD}{cell outage detection}
 \acro{CoMP}{coordinated multi-point}
 \acro{ConvLSTM}{Convolutional LSTM}
 \acro{CP}{cycle prefix}
 \acro{CPC}{cognitive pilot channel}
 \acro{CPCH}{common packet channel}
 \acro{CPE}{customer premises equipment}
 \acro{CPICH}{common pilot channel}
 \acro{CPRI}{common public radio interface}
 \acro{CPU}{central processing unit}
 \acro{CQI}{channel quality indicator}
 \acro{CR}{cognitive radio}
 \acro{CRAN}{centralized radio access network} 
 \acro{C-RAN}{cloud radio access network} 
 \acro{CRC}{cyclic redundancy check}
 \acro{CRE}{cell range expansion}
 \acro{C-RNTI}{cell radio network temporary identifier}
 \acro{CRP}{cell re-selection parameter}
 \acro{CRS}{cell-specific reference symbol}
 \acro{CRT}{cell re-selection threshold}
 \acro{CSCC}{common spectrum coordination channel}
 \acro{CSG ID}{closed subscriber group ID}
 \acro{CSG}{closed subscriber group}
 \acro{CSI}{channel state information}
 \acro{CSIR}{receiver-side channel state information}
 \acro{CSI-RS}{channel state information-reference signals}
 \acro{CSO}{cell selection offset}
 \acro{CTCH}{common traffic channel}
 \acro{CTS}{clear-to-send} 
 \acro{CU}{central unit}
 \acro{CV}{cross-validation}
 \acro{CWiND}{Centre for Wireless Network Design}
 \acro{D2D}{device to device}
 \acro{DAB}{digital audio broadcasting}
 \acro{DAC}{digital-to-analog converter}
 \acro{DAS}{distributed antenna system}
 \acro{dB}{decibel}
 \acro{dBi}{isotropic-decibel}
 %\acro{DC}{dual connectivity}
 \acro{DC}{direct current}
 \acro{DCCH}{dedicated control channel}
 \acro{DCF}{decode-and-forward}
 \acro{DCH}{dedicated channel}
 \acro{DC-HSPA}{dual-carrier high speed packet access}
 \acro{DCI}{downlink control information}
 \acro{DCM}{directional channel model}
 \acro{DCP}{dirty-paper coding}
 \acro{DCS}{digital communication system}
 \acro{DECT}{digital enhanced cordless telecommunication}
 \acro{DeNB}{donor eNodeB}
 \acro{DFP}{dynamic frequency planning}
 \acro{DFS}{dynamic frequency selection}
 \acro{DFT}{discrete Fourier transform}
 \acro{DFTS}{discrete Fourier transform spread}
 \acro{DHCP}{dynamic host control protocol}
 \acro{DL}{downlink}
 \acro{DMC}{dense multi-path components}
 \acro{DMF}{demodulate-and-forward}
 \acro{DMT}{diversity and multiplexing tradeoff}
  \acro{DNN}{deep neural network} 
 \acro{DoA}{direction-of-arrival}
 \acro{DoD}{direction-of-departure}
 \acro{DoS}{denial of service}
 \acro{DPCCH}{dedicated physical control channel}
 \acro{DPDCH}{dedicated physical data channel}
 \acro{D-QDCR}{distributed QoS-based dynamic channel reservation}
 \acro{DQL}{deep Q-learning}
  \acro{DRAN}{distributed radio access network}
 \acro{DRS}{discovery reference signal}
 \acro{DRL}{deep reinforcement learning}
 \acro{DRX}{discontinuous reception}
 \acro{DS}{down stream}
 \acro{DSA}{dynamic spectrum access}
 \acro{DSCH}{downlink shared channel}
 \acro{DSL}{digital subscriber line}
 \acro{DSLAM}{digital subscriber line access multiplexer}
 \acro{DSP}{digital signal processor}
 \acro{DT}{decision tree}
 \acro{DTCH}{dedicated traffic channel}
 \acro{DTX}{discontinuous transmission}
   \acro{DU}{distributed unit}
 \acro{DVB}{digital video broadcasting}
 \acro{DXF}{drawing interchange format}
 \acro{E2E}{end-to-end}
 \acro{EAGCH}{enhanced uplink absolute grant channel}
 \acro{EA}{evolutionary algorithm}
 \acro{EAP}{extensible authentication protocol}
 \acro{EC}{evolutionary computing}
 \acro{ECGI}{evolved cell global identifier}
 \acro{ECR}{energy consumption ratio}
 \acro{ECRM}{effective code rate map}
 \acro{EDCH}{enhanced dedicated channel}
 \acro{EE}{energy efficiency}
 \acro{EESM}{exponential effective SINR mapping}
 \acro{EF}{estimate-and-forward}
 \acro{EGC}{equal gain combining}
 \acro{EHICH}{EDCH HARQ indicator channel}
 \acro{eICIC}{enhanced intercell interference coordination}
 \acro{EIR}{equipment identity register}
 \acro{EIRP}{effective isotropic radiated power}
 \acro{ELF}{evolutionary learning of fuzzy rules}
 \acro{eMBB}{enhanced mobile broadband}
  \acro{EMR}{Electromagnetic-Radiation}
 \acro{EMS}{enhanced messaging service}
 \acro{eNB}{evolved NodeB}
 \acro{eNodeB}{evolved NodeB}
 \acro{EoA}{elevation of arrival}
 \acro{EoD}{elevation of departure}
 \acro{EPB}{equal path-loss boundary}
 \acro{EPC}{evolved packet core}
 \acro{EPDCCH}{enhanced physical downlink control channel}
 \acro{EPLMN}{equivalent PLMN}
 \acro{EPS}{evolved packet system}
 \acro{ERAB}{eUTRAN radio access bearer}
 \acro{ERGC}{enhanced uplink relative grant channel}
 \acro{ERTPS}{extended real time polling service}
 \acro{ESB}{equal downlink receive signal strength boundary}
 \acro{ESF}{even subframe}
 \acro{ESP}{encapsulating security payload}
 \acro{ETSI}{European Telecommunications Standards Institute}
 \acro{E-UTRA}{evolved UTRA}
 \acro{EU}{European Union}
 \acro{EUTRAN}{evolved UTRAN}
 \acro{EVDO}{evolution-data optimised}
 \acro{FACCH}{fast associated control channel}
 \acro{FACH}{forward access channel}
 \acro{FAP}{femtocell access point}
 \acro{FARL}{fuzzy assisted reinforcement learning}
 \acro{FCC}{Federal Communications Commission}
 \acro{FCCH}{frequency-correlation channel}
 \acro{FCFS}{first-come first-served}
 \acro{FCH}{frame control header}
 \acro{FCI}{failure cell ID}
 \acro{FD}{frequency-domain}
 \acro{FDD}{frequency division duplexing}
 \acro{FDM}{frequency division multiplexing}
 \acro{FDMA}{frequency division multiple access}
 \acro{FDTD}{finite-difference time-domain}
 \acro{FE}{front-end}
 \acro{FeMBMS}{further evolved multimedia broadcast multicast service}
 \acro{FER}{frame error rate}
 \acro{FFR}{fractional frequency reuse}
 \acro{FFRS}{fractional frequency reuse scheme}
 \acro{FFT}{fast Fourier transform}
 \acro{FFU}{flexible frequency usage}
 \acro{FGW}{femtocell gateway}
 \acro{FIFO}{first-in first-out}
 \acro{FIS}{fuzzy inference system}
 \acro{FMC}{fixed mobile convergence}
 \acro{FPC}{fractional power control}
 \acro{FPGA}{field-programmable gate array}
 \acro{FRS}{frequency reuse scheme}
 \acro{FTP}{file transfer protocol}
 \acro{FTTx}{fiber to the x}
 \acro{FUSC}{full usage of subchannels}
 \acro{GA}{genetic algorithm}
 \acro{GAN} {generic access network}
 \acro{GANC}{generic access network controller}
 \acro{GBR}{guaranteed bitrate}
 \acro{GCI}{global cell identity}
 \acro{STGCN}{Spatio-Temporal Graph convolutional network}
 \acro{GERAN}{GSM edge radio access network}
 \acro{GGSN}{gateway GPRS support node}
 \acro{GHG}{greenhouse gas}
 \acro{GMSC}{gateway mobile switching centre}
 \acro{gNB}{next generation NodeB}
 \acro{GNN}{Graph Neural Network}
 \acro{GNSS}{global navigation satellite system}
 \acro{GP}{genetic programming}
 \acro{GPON}{Gigabit passive optical network}
 \acro{GPP}{general purpose processor}
 \acro{GPRS}{general packet radio service}
 \acro{GPS}{global positioning system}
 \acro{GPU}{graphics processing unit}
 \acro{GRU}{gated recurrent unit}
 \acro{GSCM}{geometry-based stochastic channel models}
 \acro{GSM}{global system for mobile communication}
 \acro{GTD}{geometry theory of diffraction}
 \acro{GTP}{GPRS tunnel protocol}
 \acro{GTP-U}{GPRS tunnel protocol - user plane}
 %\acro{HA}{hybrid access}
 \acro{HA}{historical average}
 \acro{HARQ}{hybrid automatic repeat request}
 \acro{HBS}{home base station}
 \acro{HCN}{heterogeneous cellular network}
 \acro{HCS}{hierarchical cell structure}
  \acro{HD}{high definition}
 \acro{HDFP}{horizontal dynamic frequency planning}
 \acro{HeNB}{home eNodeB}
 \acro{HeNodeB}{home eNodeB}
 \acro{HetNet}{heterogeneous network}
 \acro{HiFi}{high fidelity}
 \acro{HII}{high interference indicator}
 \acro{HLR}{home location register}
 \acro{HNB}{home NodeB}
 \acro{HNBAP}{home NodeB application protocol}
 \acro{HNBGW}{home NodeB gateway}
 \acro{HNodeB}{home NodeB}
 \acro{HO}{handover}
 \acro{HOF}{handover failure}
 \acro{HOM}{handover hysteresis margin}
 \acro{HPBW}{half power beam width}
 \acro{HPLMN}{home PLMN}
 \acro{HPPP}{homogeneous Poison point process}
 \acro{HRD}{horizontal reflection diffraction}
 \acro{HSB}{hot spot boundary}
 \acro{HSDPA}{high speed downlink packet access}
 \acro{HSDSCH}{high-speed DSCH}
 \acro{HSPA}{high speed packet access}
 \acro{HSS}{home subscriber server}
 \acro{HSUPA}{high speed uplink packet access}
 \acro{HUA}{home user agent}
 \acro{HUE}{home user equipment}
 \acro{HVAC}{heating, ventilating, and air conditioning}
 \acro{HW}{Holt-Winters}
 \acro{IC}{interference cancellation}
 \acro{ICI}{inter-carrier interference}
 \acro{ICIC}{intercell interference coordination}
 \acro{ICNIRP}{International Commission on Non-Ionising Radiation Protection}
 \acro{ICS}{IMS centralised service}
 \acro{ICT}{information and communication technology}
 \acro{ID}{identifier}
 \acro{IDFT}{inverse discrete Fourier transform}
 \acro{IE}{information element}
 \acro{IEEE}{Institute of Electrical and Electronics Engineers}
 \acro{IETF}{Internet engineering task force}
 \acro{IFA}{Inverted-F-antennas}
 \acro{IFFT}{inverse fast Fourier transform}
 \acro{i.i.d.}{independent and identical distributed}
 \acro{IIR}{infinite impulse response}
 \acro{IKE}{Internet key exchange}
 \acro{IKEv2}{Internet key exchange version 2}
 \acro{ILP}{integer linear programming}
 \acro{IMEI}{international mobile equipment identity}
 \acro{IMS}{IP multimedia subsystem}
 \acro{IMSI}{international mobile subscriber identity}
 \acro{IMT}{international mobile telecommunications}
 \acro{INH}{indoor hotspot}
 \acro{IOI}{interference overload indicator}
 \acro{IoT}{Internet of things}
 \acro{IP}{Internet protocol}
 \acro{IPSEC}{Internet protocol security}
 \acro{IR}{incremental redundancy}
 \acro{IRC}{interference rejection combining}
 \acro{ISD}{inter site distance}
 \acro{ISI}{inter symbol interference}
 \acro{ITU}{International Telecommunication Union}
 \acro{Iub}{UMTS interface between RNC and NodeB}
 \acro{IWF}{IMS interworking function}
 \acro{JFI}{Jain's fairness index}
 \acro{KPI}{key performance indicator}
 \acro{KNN}{k-nearest neighbours}
 \acro{L1}{layer one}
 \acro{L2}{layer two}
 \acro{L3}{layer three}
 \acro{LA}{location area}
 \acro{LAA}{licensed Assisted Access}
 \acro{LAC}{location area code}
 \acro{LAI}{location area identity}
 \acro{LAU}{location area update}
 \acro{LDA}{linear discriminant analysis} 
 \acro{LIDAR}{laser imaging detection and ranging}
 \acro{LLR}{log-likelihood ratio}
 \acro{LLS}{link-level simulation}
 \acro{LMDS}{local multipoint distribution service}
 \acro{LMMSE}{linear minimum mean-square-error}
 \acro{LoS}{line-of-sight}
 \acro{LPC}{logical PDCCH candidate}
 \acro{LPN}{low power node}
 \acro{LR}{likelihood ratio}
 \acro{LSAS}{large-scale antenna system}
 \acro{LSP}{large-scale parameter}
 \acro{LSTM}{long short term memory cell}
 \acro{LTE/SAE}{long term evolution/system architecture evolution}
 \acro{LTE}{long term evolution}
 \acro{LTE-A}{long term evolution advanced}
 \acro{LUT}{look up table}
 \acro{MAC}{medium access control}
 \acro{MaCe}{macro cell}
  \acro{MAE}{mean absolute error}
 \acro{MAP}{media access protocol}
 \acro{MAPE}{mean absolute percentage error}
 \acro{MAXI}{maximum insertion}
 \acro{MAXR}{maximum removal}
 \acro{MBMS}{multicast broadcast multimedia service} % <<<
 \acro{MBS}{macrocell base station}
 \acro{MBSFN}{multicast-broadcast single-frequency network}
 \acro{MC}{modulation and coding}
 \acro{MCB}{main circuit board}
 \acro{MCM}{multi-carrier modulation}
 \acro{MCP}{multi-cell processing}
 \acro{MCPA}{multi-carrier power amplifier}
 \acro{MCS}{modulation and coding scheme}
 \acro{MCSR}{multi-carrier soft reuse}
 \acro{MDAF}{management data analytics function}
 \acro{MDP}{markov decision process }
 \acro{MDT}{minimisation of drive tests}
 \acro{MEA}{multi-element antenna}
 \acro{MeNodeB}{Master eNodeB}
 \acro{MGW}{media gateway}
 \acro{MIB}{master information block}
 \acro{MIC}{mean instantaneous capacity}
 \acro{MIESM}{mutual information effective SINR mapping}
 \acro{MIMO}{multiple-input multiple-output}
 \acro{MINI}{minimum insertion}
 \acro{MINR}{minimum removal}
 \acro{MIP}{mixed integer program}
 \acro{MISO}{multiple-input single-output}
 \acro{ML}{machine learning}
 \acro{MLB}{mobility load balancing}
 \acro{MLB}{mobility load balancing}
 \acro{MM}{mobility management}
 \acro{MME}{mobility management entity}
 \acro{mMIMO}{massive multiple-input multiple-output}
 \acro{MMSE}{minimum mean square error}
 \acro{mMTC}{massive machine type communication}
 \acro{MNC}{mobile network code}
 \acro{MNO}{mobile network operator}
 \acro{MOS}{mean opinion score}
 \acro{MPC}{multi-path component}
 \acro{MR}{measurement report}
 \acro{MRC}{maximal ratio combining}
 \acro{MR-FDPF}{multi-resolution frequency-domain parflow}
 \acro{MRO}{mobility robustness optimisation}
 \acro{MRT}{Maximum Ratio Transmission}
 \acro{MS}{mobile station}
 \acro{MSC}{mobile switching centre}
 \acro{MSE}{mean square error}
 \acro{MSISDN}{mobile subscriber integrated services digital network number}
 \acro{MUE}{macrocell user equipment}
 \acro{MU-MIMO}{multi-user MIMO}
 \acro{MVNO}{mobile virtual network operators}
 \acro{NACK}{negative acknowledgment}
 \acro{NAS}{non access stratum}
 \acro{NAV}{network allocation vector}
 \acro{NB}{Naive Bayes}   
 \acro{NCL}{neighbour cell list}
 \acro{NEE}{network energy efficiency}
  \acro{NF}{Network Function}
 \acro{NFV}{Network Functions Virtualization}
 \acro{NG}{next generation}
 \acro{NGMN}{next generation mobile networks}
 \acro{NG-RAN}{next generation radio access network} 
 \acro{NIR}{non ionisation radiation}
 \acro{NLoS}{non-line-of-sight}
 \acro{NN}{nearest neighbour} 
 \acro{NR}{new radio}
 \acro{NRTPS}{non-real-time polling service}
 \acro{NSS}{network switching subsystem}
 \acro{NTP}{network time protocol}
 \acro{NWG}{network working group}
 \acro{NWDAF}{network data analytics function} 
 \acro{OA}{open access}
 \acro{OAM}{operation, administration and maintenance}
 \acro{OC}{optimum combining}
 \acro{OCXO}{oven controlled oscillator}
 \acro{ODA}{omdi-directional antenna} 
 \acro{ODU}{optical distribution unit}
 \acro{OFDM}{orthogonal frequency division multiplexing}
 \acro{OFDMA}{orthogonal frequency division multiple access}
 \acro{OFS}{orthogonally-filled subframe}
 \acro{OLT}{optical line termination}
 \acro{ONT}{optical network terminal}
 \acro{OPEX}{operational expenditure}
 \acro{OSF}{odd subframe}
 \acro{OSI}{open systems interconnection}
 \acro{OSS}{operation support subsystem}
 \acro{OTT}{over the top}
 \acro{P2MP}{point to multi-point}
 \acro{P2P}{point to point}
 \acro{PAPR}{peak-to-average power ratio}
 \acro{PA}{power amplifier}
 \acro{PBCH}{physical broadcast channel}
 \acro{PC}{power control}
 \acro{PCB}{printed circuit board}
 \acro{PCC}{primary carrier component}
 \acro{PCCH}{paging control channel}
 \acro{PCCPCH}{primary common control physical channel}
 \acro{PCell}{primary cell}
 \acro{PCFICH}{physical control format indicator channel}
 \acro{PCH}{paging channel}
 \acro{PCI}{physical layer cell identity}
 \acro{PCPICH}{primary common pilot channel}
 \acro{PCPPH}{physical common packet channel}
 \acro{PDCCH}{physical downlink control channel}
 \acro{PDCP}{packet data convergence protocol}
 \acro{PDF}{probability density function}
 \acro{PDSCH}{physical downlink shared channel}
 \acro{PDU}{packet data unit}
 \acro{PeNB}{pico eNodeB}
 \acro{PeNodeB}{pico eNodeB}
 \acro{PF}{proportional fair}
 \acro{PGW}{packet data network gateway}
 \acro{PGFL}{probability generating functional}
 \acro{PhD}{doctor of philosophy}
 \acro{PHICH}{physical HARQ indicator channel}
 \acro{PHY}{physical}
 \acro{PIC}{parallel interference cancellation}
 \acro{PKI}{public key infrastructure}
 \acro{PL}{path loss}
 \acro{PMI}{precoding-matrix indicator}
 \acro{PLMN ID}{public land mobile network identity}
 \acro{PLMN}{public land mobile network}
 \acro{PML}{perfectly matched layer}
 \acro{PMF}{probability mass function}
 \acro{PMP}{point to multi-point}
 \acro{PN}{pseudorandom noise}
 \acro{POI}{point of interest}
 \acro{PON}{passive optical network}
 \acro{POP}{point of presence}
 \acro{PP}{point process}
 \acro{PPP}{Poisson point process}
 \acro{PPT}{PCI planning tools}
 \acro{PRACH}{physical random access channel}
 \acro{PRB}{physical resource block}
 \acro{PSC}{primary scrambling code}
 \acro{PSD}{power spectral density}
 \acro{PSS}{primary synchronisation channel}
 \acro{PSTN}{public switched telephone network}
 \acro{PTP}{point to point}
 \acro{PUCCH}{Physical Uplink Control Channel}
 \acro{PUE}{picocell user equipment}
 \acro{PUSC}{partial usage of subchannels}
 \acro{PUSCH}{physical uplink shared channel}
 \acro{QAM}{quadrature amplitude modulation}
 \acro{QCI}{QoS class identifier}
 \acro{QoE}{quality of experience}
 \acro{QoS}{quality of service}
 \acro{QPSK}{quadrature phase-shift keying}
 \acro{RAB}{radio access bearer}
 \acro{RACH}{random access channel}
 \acro{RADIUS}{remote authentication dial-in user services}
 \acro{RAN}{radio access network}
 \acro{RANAP}{radio access network application part}
 \acro{RAT}{radio access technology}
 \acro{RAU}{remote antenna unit}
 \acro{RAXN}{relay-aided x network}
 \acro{RB}{resource block}
 \acro{RCI}{re-establish cell id}
 \acro{RE}{resource efficiency}
 \acro{REB}{range expansion bias}
 \acro{REG}{resource element group}
 \acro{RF}{radio frequency}
  \acro{RFID}{radio frequency identification}
 \acro{RFP}{radio frequency planning}
 \acro{RI}{rank indicator}
 \acro{RL}{reinforcement learning}
 \acro{RLC}{radio link control}
 \acro{RLF}{radio link failure}
 \acro{RLM}{radio link monitoring}
 \acro{RMA}{rural macrocell}
 \acro{RMS}{root mean square}
 \acro{RMSE}{root mean square error}
 \acro{RN}{relay node}
 \acro{RNC}{radio network controller}
 \acro{RNL}{radio network layer}
 \acro{RNN}{recurrent neural network}
 \acro{RNP}{radio network planning}
 \acro{RNS}{radio network subsystem}
 \acro{RNTI}{radio network temporary identifier}
 \acro{RNTP}{relative narrowband transmit power}
 \acro{RPLMN}{registered PLMN}
 \acro{RPSF}{reduced-power subframes}
 \acro{RR}{round robin}
 \acro{RRC}{radio resource control}
 \acro{RRH}{remote radio head}
 \acro{RRM}{radio resource management}
 \acro{RS}{reference signal}
 \acro{RSC}{recursive systematic convolutional}
 \acro{RS-CS}{resource-specific cell-selection}
 \acro{RSQ}{reference signal quality}
 \acro{RSRP}{reference signal received power}
 \acro{RSRQ}{reference signal received quality}
 \acro{RSS}{reference signal strength}
 \acro{RSSI}{receive signal strength indicator}
 \acro{RTP}{real time transport}
 \acro{RTPS}{real-time polling service}
 \acro{RTS}{request-to-send}
 \acro{RTT}{round trip time}
  \acro{RU}{remote unit}
  \acro{RV}{random variable}
 \acro{RX}{receive}
 \acro{S1-AP}{s1 application protocol}
 \acro{S1-MME}{s1 for the control plane}
 \acro{S1-U}{s1 for the user plane}
 \acro{SA}{simulated annealing}
 \acro{SACCH}{slow associated control channel}
 \acro{SAE}{system architecture evolution}
 \acro{SAEGW}{system architecture evolution gateway}
 \acro{SAIC}{single antenna interference cancellation}
 \acro{SAP}{service access point}
 \acro{SAR}{specific absorption rate}
 \acro{SARIMA}{seasonal autoregressive integrated moving average}
 \acro{SAS}{spectrum allocation server}
 \acro{SBS}{super base station}
 \acro{SCC}{standards coordinating committee}
 \acro{SCCPCH}{secondary common control physical channel}
 \acro{SCell}{secondary cell}
 \acro{SCFDMA}{single carrier FDMA}
 \acro{SCH}{synchronisation channel}
 \acro{SCM}{spatial channel model}
 \acro{SCN}{small cell network}
 \acro{SCOFDM}{single carrier orthogonal frequency division multiplexing}
 \acro{SCP}{single cell processing}
 \acro{SCTP}{stream control transmission protocol}
 \acro{SDCCH}{standalone dedicated control channel}
 \acro{SDMA}{space-division multiple-access}
  \acro{SDO}{standard development organization}
 \acro{SDR}{software defined radio}
 \acro{SDU}{service data unit}
 \acro{SE}{spectral efficiency}
 \acro{SeNodeB}{secondary eNodeB}
 \acro{Seq2Seq}{Sequence-to-sequence}
 \acro{SFBC}{space frequency block coding}
 \acro{SFID}{service flow ID}
 \acro{SG}{signalling gateway}
 \acro{SGSN}{serving GPRS support node}
 \acro{SGW}{serving gateway}
 \acro{SHAP}{SHapley Additive exPlanations}
 \acro{SI}{system information}
 \acro{SIB}{system information block}
 \acro{SIB1}{systeminformationblocktype1}
 \acro{SIB4}{systeminformationblocktype4}
 \acro{SIC}{successive interference cancellation}
 \acro{SIGTRAN}{signalling transport}
 \acro{SIM}{subscriber identity module}
 \acro{SIMO}{single input multiple output}
 \acro{SINR}{signal to interference plus noise ratio}
 \acro{SIP}{session initiated protocol}
 \acro{SIR}{signal to interference ratio}
 \acro{SISO}{single input single output}
 \acro{SLAC}{stochastic local area channel}
 \acro{SLL}{secondary lobe level}
 \acro{SLNR}{signal to leakage interference and noise ratio}
 \acro{SLS}{system-level simulation}
 \acro{SMAPE}{symmetric mean absolute percentage error}
 \acro{SMB}{small and medium-sized businesses}
 \acro{SmCe}{small cell}
 \acro{SMS}{short message service}
 \acro{SN}{serial number}
 \acro{SNMP}{simple network management protocol}
 \acro{SNR}{signal to noise ratio}
 \acro{SOCP}{second-order cone programming}
 \acro{SOHO}{small office/home office}
 \acro{SON}{self-organising network}
 \acro{son}{self-organising networks}
 \acro{SOT}{saving of transmissions}
 \acro{SPS}{spectrum policy server}
 \acro{SRS}{sounding reference signals}
 \acro{SS}{synchronization signal}
 \acro{SSL}{secure socket layer}
 \acro{SSMA}{spread spectrum multiple access}
 \acro{SSS}{secondary synchronisation channel}
 \acro{ST}{spatio temporal}
 \acro{STA}{steepest ascent}
 \acro{STBC}{space-time block coding}
 \acro{SUI}{stanford university interim}
 \acro{SVR}{support vector regression}
 \acro{TA}{timing advance}
 \acro{TAC}{tracking area code}
 \acro{TAI}{tracking area identity}
 \acro{TAS}{transmit antenna selection}
 \acro{TAU}{tracking area update}
 \acro{TCH}{traffic channel}
 \acro{TCO}{total cost of ownership}
 \acro{TCP}{transmission control protocol}
 \acro{TCXO}{temperature controlled oscillator}
 \acro{TD}{temporal difference}
 \acro{TDD}{time division duplexing}
 \acro{TDM}{time division multiplexing}
 \acro{TDMA}{time division multiple access}
  \acro{TDoA}{time difference of arrival}
 \acro{TEID}{tunnel endpoint identifier}
 \acro{TLS}{transport layer security}
 \acro{TNL}{transport network layer}
  \acro{ToA}{time of arrival}
 \acro{TP}{throughput}
 \acro{TPC}{transmit power control}
 \acro{TPM}{trusted platform module}
 \acro{TR}{transition region}
  \acro{TRX}{transceiver}
 \acro{TS}{tabu search}
 \acro{TSG}{technical specification group}
 \acro{TTG}{transmit/receive transition gap}
 \acro{TTI}{transmission time interval}
 \acro{TTT}{time-to-trigger}
 \acro{TU}{typical urban}
 \acro{TV}{television}
 \acro{TWXN}{two-way exchange network}
 \acro{TX}{transmit}
 \acro{UARFCN}{UTRA absolute radio frequency channel number}
 \acro{UAV}{unmanned aerial vehicle}
 \acro{UCI}{uplink control information}
 \acro{UDP}{user datagram protocol}
 \acro{UDN}{ultra-dense network}
 \acro{UE}{user equipment}
 \acro{UGS}{unsolicited grant service}
 \acro{UICC}{universal integrated circuit card}
 \acro{UK}{united kingdom}
 \acro{UL}{uplink}
 \acro{UMA}{unlicensed mobile access}
 \acro{UMi}{urban micro}
 \acro{UMTS}{universal mobile telecommunication system}
 \acro{UN}{United Nations}
 \acro{URLLC}{ultra-reliable low-latency communication}
 \acro{US}{upstream}
 \acro{USIM}{universal subscriber identity module}
 \acro{UTD}{theory of diffraction}
 \acro{UTRA}{UMTS terrestrial radio access}
 \acro{UTRAN}{UMTS terrestrial radio access network}
 \acro{UWB}{ultra wide band}
 \acro{VD}{vertical diffraction}
 \acro{VDFP}{vertical dynamic frequency planning}
 \acro{VDSL}{very-high-bit-rate digital subscriber line}
 \acro{VeNB}{virtual eNB}
 \acro{VeNodeB}{virtual eNodeB}
 \acro{VIC}{victim cell}
 \acro{VLR}{visitor location register}
 \acro{VNF}{virtual network function}
 \acro{VoIP}{voice over IP}
 \acro{VoLTE}{voice over LTE}
 \acro{VPLMN}{visited PLMN}
 \acro{VR}{visibility region}
  \acro{VRAN}{virtualized radio access network}
 \acro{WCDMA}{wideband code division multiple access}
 \acro{WEP}{wired equivalent privacy}
 \acro{WG}{working group}
 \acro{WHO}{world health organisation}
 \acro{Wi-Fi}{Wi-Fi}
 \acro{WiMAX}{wireless interoperability for microwave access}
 \acro{WiSE}{wireless system engineering}
 \acro{WLAN}{wireless local area network}
 \acro{WMAN}{wireless metropolitan area network}
 \acro{WNC}{wireless network coding}
 \acro{WRAN}{wireless regional area network}
 \acro{WSEE}{weighted sum of the energy efficiencies}
 \acro{WPEE}{weighted product of the energy efficiencies}
 \acro{WMEE}{weighted minimum of the energy efficiencies}
 \acro{X2}{x2}
 \acro{X2-AP}{x2 application protocol}
 \acro{ZF}{zero forcing}

 \end{acronym}

\end{document}